\documentclass[twocolumn]{aastex631}

\makeatletter
\DeclareRobustCommand{\HI}{%
  \mbox{H\check@mathfonts\fontsize\sf@size\z@\selectfont I}%
}
\makeatother

\makeatletter
\DeclareRobustCommand{\HII}{%
  \mbox{H\check@mathfonts\fontsize\sf@size\z@\selectfont II}%
}
\makeatother

\usepackage{lipsum}
\usepackage{multirow}
\usepackage{amsmath}
\usepackage{natbib}
\usepackage{placeins}
\usepackage{float}

\begin{document}

\title{A Pre-Infall Magellanic Analog-- Corona and Stream Formation in the \textsc{HESTIA} Cosmological Simulations}

\author[0009-0002-9263-5574]{Robin Chisholm} 
\affiliation{University of Wisconsin, Madison, Department of Physics, 1150 University Avenue, Madison, WI 53706, USA}

\author[0000-0003-2676-8344]{Elena D'Onghia} 
\affiliation{University of Wisconsin, Madison, Department of Physics, 1150 University Avenue, Madison, WI 53706, USA}
\affiliation{University of Wisconsin, Madison, Department of Astronomy, 475 N. Charter Street, Madison, WI 53706, USA}
\affiliation{INAF - Osservatorio Astrofisico di Torino, via Osservatorio 20, 10025 Pino Torinese (TO), Italy}

\author[0000-0002-6406-0016]{Noam Libeskind}
\affiliation{Leibniz-Institut f\"ur Astrophysik Potsdam (AIP), An der Sternwarte 16, D-14482 Potsdam, Germany}

\author[0000-0001-9982-0241]{Scott Lucchini}
\affiliation{Harvard \& Smithsonian, Center for Astrophysics, 60 Garden Street, Cambridge, MA 02138, USA}

\author[0000-0003-0724-4115]{Andrew J. Fox}
\affil{AURA for ESA, Space Telescope Science Institute, 3700 San Martin Drive, Baltimore, MD 21218}

\author[0000-0001-6516-7459]{Matthias Steinmetz}
\affiliation{Leibniz-Institut f\"ur Astrophysik Potsdam (AIP), An der Sternwarte 16, D-14482 Potsdam, Germany}

\begin{abstract}
We identify and investigate a pre-infall analog of the Large and Small Magellanic Clouds (LMC, SMC) in the \textsc{Hestia} suite of constrained cosmological simulations. The system, dynamically isolated from the Local Group, evolves over $\sim$6 Gyr and forms a multiphase warm coronal halo and a neutral gas stream via repeated tidal interactions, $\sim150$ kpc in length. 
The LMC-analog’s corona forms self-consistently through virial accretion and inhibits the survival of clumpy neutral structures beyond $\sim$600 Myr. 
The SMC analog remains bound through to $z=0$, and the pair also exhibits bridge-like and leading-arm features. These results suggest that while most of the ionized Stream is formed by the LMC coronal gas, the neutral gas stream, bridge, and leading arm components of the Magellanic System can arise from dwarf–dwarf interactions prior to infall, while the survival and ionization of these features likely require additional environmental processing. 
Furthermore, we identify a stellar component out of phase to the neutral component of the stream, implying that if the Magellanic stellar stream exists, it may not be spatially coexistent to the dominant \ion{H}{1} stream.
This system offers a valuable pre-infall reference point for interpreting the Magellanic System and identifying analogs beyond the Local Group.

\end{abstract}




\section{Introduction} \label{sec:intro}

The Magellanic System, consisting of the Large and Small Magellanic Clouds (LMC and SMC) as well as a variety of associated gaseous structures, represents a compelling example of an interacting dwarf galaxy pair in the Local Group \citep[see][]{donghia+2016}. 
This system is dynamically complex, with a prominent gaseous stream, the Magellanic Stream, that extends $200^{\circ}$ across the sky \citep{nidever+2010}. Recent mass estimates suggest that the LMC is significantly more massive than previously thought, with a halo mass of the order of $10^{11.3}-10^{11.5} M_{\odot}$ 
\citep[e.g.,][]{erkal+2019, patel+2020, besla+2007, penarrubia+2016, cavieres+2025, shipp+2021, watkins+2024}. 
These revised estimates raise new questions about the circumgalactic medium (CGM) of the LMC, including the role of a warm coronal gas envelope that may have influenced the evolution of the Magellanic Stream \citep{Lucchini+2020, Lucchini+2021, Lucchini+2024, Krishnarao+2022}; the existence of such a coronal gas halo around the LMC has been supported both theoretically and observationally.  

Recent ultraviolet absorption line studies have revealed an extended, ionized gas envelope around the LMC, extending up to 35 kpc from its disk \citep{Krishnarao+2022, sapna+2024}. These studies confirmed that this gas is kinematically consistent with an LMC-associated truncated corona and the ionized component of the Magellanic Stream rather than being part of the Milky Way's CGM, indicating that massive dwarf galaxies can sustain their own gaseous halos even in the presence of a larger host galaxy. This result is consistent with previous studies that indicate that massive dwarf halos, such as the LMC, should host warm virialized coronal gas \citep[e.g.,][]{nuza+2014, jahn+2022}. More recent high-resolution hydrodynamical simulations have shown that the presence of such a corona can significantly alter the evolution of tidally stripped gas, potentially influencing the formation of the Magellanic Stream \citep{Lucchini+2020, Lucchini+2024}. However, most previous numerical studies of the Magellanic system have focused on its interaction with the Milky Way \citep[e.g.,][]{moore+2004, mastropietro+2005} or modeled the LMC and SMC in isolation \citep{Besla+2012, pardy+2018}, without capturing the self-consistent formation and evolution of a multiphase CGM.

These findings raise key questions regarding the properties and evolution of the LMC corona prior to its accretion by the Milky Way; specifically, the role of this circumgalactic gas in shielding the Magellanic Stream from ionization and regulating gas accretion onto the LMC remains an open issue. Moreover, a cosmological context is essential to naturally form a stable, multiphase coronal structure through virialization, avoiding the need for fine-tuned initial conditions. This enables a more realistic assessment of the role of CGM in shaping the gas dynamics of the system.

In this work, we leverage the \textsc{Hestia} (High-Resolution Environmental Simulations of the Immediate Area) cosmological simulation suite to study the CGM of a Magellanic-analog system within a full cosmological context. These simulations offer a unique opportunity to investigate the formation and properties of a warm coronal envelope in an isolated dwarf galaxy pair before its accretion onto a Milky Way-mass host \citep{donghia+2008}. Unlike previous studies, our analysis identifies a resolved warm coronal envelope in a Magellanic system-analog within a cosmological volume simulation, specifically in a constrained realization of the Local Group. While similar structures have been reported in zoom-in simulations \citep{jahn+2022}, this is the first identification of such a system including both the LMC- and SMC-analog pair in a cosmological setting, allowing investigation of their CGM and tidal features prior to infall.


This analog features a resolved neutral gas stream (with trailing and leading components) formed through dwarf–dwarf interaction and allows us to study the evolution of such a structure embedded in a realistic hot CGM. By tracking the stream's morphology, density, and mass over more than 1.5 Gyr, we provide new theoretical insight into the survivability of stripped neutral gas. This analog provides a rare, self-consistent framework for studying how dwarf–dwarf tidal interactions shape gaseous structures in realistic halos, and allows us to test the survivability of neutral gas streams in a cosmological setting — a key theoretical and observational question for the Magellanic System and beyond.

This paper is structured as follows. In \S\ref{sec:methods}, we describe the \textsc{Hestia} simulations and methodology to identify different components of the CGM and other gaseous structures. In \S\ref{sec:results} we describe in detail the properties of the Magellanic-analog system, including the characterization of the coronal gas, the formation of an \ion{H}{1}-rich stream, and the ubiquity of coronae in massive dwarfs within the simulation suite. In \S\ref{sec:discussion}, we discuss the implications of our findings. And finally, in \S\ref{sec:conclusion}, we provide concluding remarks.

\begin{figure*}[htp!]
\centering
\includegraphics[width=0.95\linewidth]{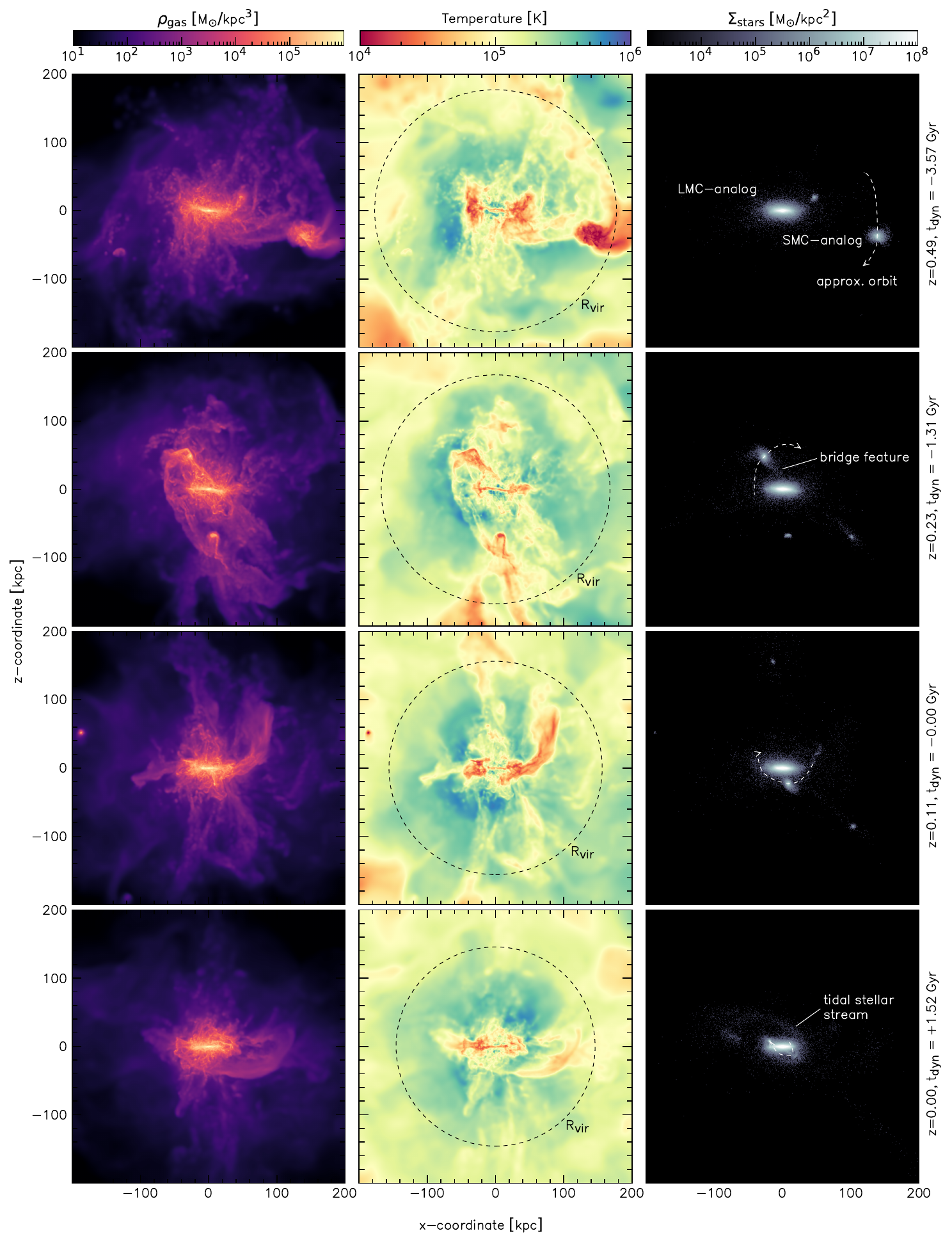}
\caption{Edge-on projections of the simulated Magellanic-analog system at various snapshots, illustrating the evolution of the interacting dwarf galaxy pair.}
\label{fig:images}
\end{figure*}

\section{Methodology} \label{sec:methods}


\subsection{The \textsc{HESTIA} simulations}

The \textsc{Hestia} project consists of a suite of constrained cosmological zoomed-in simulations designed to reproduce the local universe with high fidelity \citep{Libeskind+2020}. Low resolution, dark-matter only simulations are run with initial conditions constrained by the observed peculiar velocity of nearby galaxies, groups, and clusters to accurately simulate the local cosmography. These constraints operate on scales of several Mpc and encode the large-scale tidal field of the local universe, while allowing the small-scale structure (e.g., dwarf galaxies and their interactions) to arise stochastically within that tidal environment.


Halo pairs that resemble the Local Group are identified and then selected for higher-resolution magneto-hydrodynamical simulation runs using \textsc{Arepo} \citep{springel+2010, weinberger+2020}, with baryonic physics following the \textsc{Auriga} galaxy formation model \citep{grand+2017}. This includes prescriptions for star formation, supernova \& AGN feedback, metal enrichment, and radiative cooling. Post-processing of the simulations output for halo/subhalo/galaxy identification was performed by the \textit{Amiga} Halo Finder. Post-processing includes the identification of halos, subhalos, and galaxies using AHF, allowing us to trace both the dark matter and baryonic evolution of each system.\footnote{\textsc{Ahf} is publicly available at \url{http://popia.ft.uam.es/AHF/index.html}}, with bulk baryonic properties computed, as well as merger trees and halo growth data at each redshift \citep{knollmann+2009}.\footnote{In addition to \textsc{Ahf}, friends-of-friends (FOF) catalogs were used, exclusively for Black Hole seeding in the galaxy formation model.} The simulations adopt a Planck 2014 cosmology with parameters: $\sigma_8=0.83$, $H_0 = 100h$ km s$^{-1}$ Mpc$^{-1}$ where $h=0.677$, $\Omega_M = 0.270$ and $\Omega_b = 0.048$, see \citet{Libeskind+2020}.

The zoom-in regions of the three highest resolution runs are populated with $8192^3$ effective particles within two overlapping $3.7$ Mpc spheres centered on either main $z=0$ LG member. A mass resolution of $\sim 2\times10^4$ M$_{\odot}$ and a spatial resolution of $\sim 220$ pc for the gas cells is achieved, allowing us to resolve detailed circumgalactic structures such as the warm corona and tidally stripped neutral gas at unprecedented precision within the appropriate cosmographic environment. Furthermore, due to the relatively large zoom-in region, dozens of ``field" dwarfs are simultaneously simulated alongside the main LG members and their dwarf satellites.

Within one of the highest-resolution runs, we identify one of these field dwarf-systems: an interacting dwarf galaxy pair consisting of a larger host with $M_{\text{host}}\sim 10^{11.5}$ M$_{\odot}$ and a smaller companion with mass $M_{\text{satellite}}\sim 10^{10}$ M$_{\odot}$. This system is dynamically isolated at $z=0$, located at a distance $d \gtrsim 1$ Mpc from both the Milky Way- and M31- analog halos and produces tidally-stripped neutral gaseous structures, making it an ideal candidate for studying the pre-infall phase of a Magellanic-analog system.

\subsection{Identifying and characterizing the system components}\label{sec:corona_characterization}
To analyze the circumgalactic medium of massive dwarfs and extract cool gaseous structures, we classify gas into distinct physical phases based solely on temperature thresholds, following CGM literature \citep[e.g.,][]{tumlinson+2017, stern+2021}. Other quantities such as density, metallicity, and neutral fraction are tracked during the simulation but are not used as classification criteria. All chemical abundances and neutral fractions are computed on-the-fly during the simulation run.

\begin{itemize}
\item \textbf{Warm CGM (Coronal Gas):} 
This phase includes gas with temperatures $T > 10^5$ K; typical densities range from $n_H\sim$  10$^{-6}$ to 10$^{-4}$ cm$^{-3}$, but density is not a selection criterion.
This phase corresponds to the predicted virialized corona in a massive dwarf halo \citep{Lucchini+2020}\footnote{Functionally no gas lies above $10^6$ K surrounding these massive dwarfs, and so there is no hot CGM phase present, as expected from the virial theorem.}.

\item \textbf{Cool CGM:} This category is used descriptively to refer to gas with $T < 10^5$ K, which includes filamentary accretion flows \citep{keres2007} and warm–cool CGM interfaces. We do not analyze this component further in the current work.

\item \textbf{Neutral Gas Stream:} 
Gas cells associated with the stream and leading arm are selected based on angular momentum and kinematics, identifying material stripped from the SMC-analog. The neutral hydrogen fraction $f_{\text{HI}}$, computed on-the-fly, is used to weight the gas in visualizations and mass calculations but is not used for selection.

\end{itemize}

To avoid contamination from the gaseous disk and unrelated halos, a gas cell is considered part of the CGM only if it lies within a spherical shell defined by $R_{\rm vir}/10<r<R_{\rm vir}$ from the center of the halo. To quantify the properties of the corona, we analyze the gas in $T\text{-}n_{\rm H}$ phase space. The warm and cool CGM phases occupy characteristic regions in this space, with their distributions shaped by halo mass and environmental context. For the warm CGM in particular, we isolate this component and fit a two-dimensional Gaussian in log-temperature and log-density space to derive its mean properties and associated dispersions. This methodology allows us to consistently define the extent and structure of coronal gas across varying halo masses and to assess interactions with neutral tidal features. Unless otherwise specified, quoted uncertainties represent $1\sigma$ standard deviations from the Gaussian fit.


For our analysis of coronae across the full sample of massive dwarfs in the Hestia suite, we require each galaxy to host $>100$ warm-phase CGM gas cells. This corresponds to a minimum ionized CGM mass of $\sim2\times10^6$ M$_{\odot}$, providing a conservative detection threshold based on the resolution limits of our simulations.



\subsection{Dynamical Time}
\label{sec:dynamical_time}

To enable a direct comparison between the orbital history of our Magellanic analog system and that of previous models of LMC–SMC interactions, we define a dynamical time, $t_{\rm dyn}$, based on the local orbital dynamics of the interacting dwarf galaxies. The zero point of this time coordinate, $t_{\rm dyn}(0\text{ Gyr})$, corresponds to the \textit{stopping time} adopted in \citet{Besla+2012} and \citet{Lucchini+2020}, i.e. the time at which the isolated LMC-SMC interaction is stopped so that they can be placed a the edge of the Milky Way's potential. 
We define this “zero dynamical time” by aligning the successive pericenters of the orbits in our Magellanic-analog system with those in the second, isolated model from \citet{Besla+2012}, which excludes the influence of the Milky Way. This alignment is illustrated in Figure 2, upper left panel.
Therefore, we can directly compare the local dynamics of our LMC-SMC pair to the \textit{pre-infall} state of the LMC-SMC pairs in previous models,
as well as the dynamics of the pair for $\sim1.5$ Gyr after the stopping time --roughly the estimated amount of time it take the Clouds to travel from the edge of the Milky Way halo to their current position-- without the influence of a Milky Way potential.
This stopping time also marks the epoch when the majority of neutral gas is removed from the SMC, approximately at its third pericentric passage. 
In our simulation, this zero dynamical time occurs at redshift $z \sim 0.11$, or equivalently at a lookback time of $\sim$1.52 Gyr.


\subsection{Coordinate Transformations and Column Densities}
\label{sec:coordinate_transformations}

The static reference frame of the massive dwarf is defined such that the stellar angular momentum within 10\% of the virial radius, $\vec{L}_{\star}(<R_{\rm vir}/10)$, points in the $+z$-direction. Thus, the disk of the massive dwarf lies in the $(x,y)$-plane. In the case of our Magellanic-analog system, the smaller dwarf's orbital plane corresponds roughly to the $(x,z)$-plane.

To compare with observations of the \ion{H}{1} Magellanic Stream, we construct column density ($N_{\text{HI}}$) maps by placing a fictitious observer 55 kpc above the disk plane of the massive dwarf, i.e., at position $\vec{r}_{\rm{observer}}=(0,0,+55)$ kpc, following distance estimates from \citet{bruns+2005}. We then project the three-dimensional gas distribution in the massive dwarf's frame onto a two-dimensional spherical surface, compressing radial vectors into lines of sight. The columns' densities are calculated by integrating the volumetric \ion{H}{1} density, $n_{\text{HI}}$, along each line of sight.

This projection simulates the viewpoint of an observer looking directly down onto the LMC analog disk from above (i.e. in the negative z-direction). The resulting column density maps are shown in the upper right and lower panels of Figure 2. These stereographic projections reveal the spatial distribution and evolution of the neutral gas stream and bridge-like structures from the perspective of this fixed observer, allowing visual comparisons with observational \ion{H}{1} maps of the Magellanic System.

\section{Results}\label{sec:results}

\subsection{Identification of a Magellanic-Analog System}\label{sec:indentification_of_system}
Within the \textsc{Hestia} cosmological simulation suite, we visually identify an interacting field dwarf galaxy pair that serves as a compelling analog to the Magellanic System prior to its accretion by the Milky Way. 

The system includes a massive dwarf galaxy, hereafter referred to as the ``LMC-analog,'' with a halo mass of $M_{\rm halo} =3.02\times 10^{11}$ M$_{\odot}$ and a stellar mass of $M_{*} = 5.75 \times10^{9}$ M$_{\odot}$, placing it slightly above the upper range of observational estimates for the LMC \citep{cavieres+2025}. The galaxy also hosts a central massive black hole with $M_{\rm bh} = 1.23 \times 10^7$ M$_{\odot}$, which exceeds current mass estimates from hypervelocity star constraints \citep{han+2025}, but remains within the upper limits derived from stellar rotation fields \citep{boyce+2017}.

Its companion, hereafter the ``SMC-analog,'' has a halo mass of $M_{\rm halo} = 1.26\times10^{10}\text{ M}_{\odot}$, and a stellar mass, $M_{*} = 3.80\times10^{8}\text{ M}_{\odot}$, consistent with values inferred for the pre-infall SMC \citep{bekki+2009, snez+2004}. At $z=0$, the pair is located $\gtrsim1$ Mpc from both the Milky Way- and M31-analogs, making it dynamically isolated from any major host potential. This isolation allows us to study the evolution of the system in a pre-infall state.

The two dwarfs orbit each other for $\sim$6 Gyr, completing multiple passages. The SMC-analog completes approximately four pericentric passages around the LMC-analog during this period, with orbital separations gradually decreasing due to tidal interaction and hydrodynamical drag. The late pericenter distances range from $\sim$45 to 15 kpc, and align well with the orbital configurations explored in \citet{Besla+2012} and \citet{Lucchini+2020}. During this interaction, tidal forces strip a significant amount of gas from the SMC analog, forming a coherent neutral hydrogen stream that extends over approximately 150 kpc. The resulting \ion{H}{1} mass of the stream is $M_{\rm{HI}} = 2.09 \times10^8 \text{ M}_{\odot}$, consistent with estimates of the observed Magellanic Stream \citep{bruns+2005, fox+2014}. 
Despite the strength of the tidal interaction, the SMC-analog does not fully merge with the LMC-analog during the simulation. It retains a distinct dark matter and stellar halo through to $z=0$, as confirmed by its separate kinematic identity and spatial coherence. This persistent separation is visible in Figure \ref{fig:images} (\textit{right panels}), where the SMC-analog remains dynamically bound and distinct. Therefore, the formation of the neutral gas stream in our model rises from repeated close passages—consistent with the observed state of the Magellanic Clouds, where the SMC has not yet merged with the LMC.

Figure~\ref{fig:images} shows the evolution of the system through edge-on projections of three key quantities: total volumetric gas density, $\rho_{\rm gas}$ (\textit{left column}), binned gas temperature (\textit{center column}), and stellar surface density, $\Sigma_{\rm stars}$ (\textit{right column}). The dashed circles in the center panels denote the virial radius of the LMC-analog at each redshift. The LMC- and SMC-analogs are labeled in the upper right panel, and the orbital paths of the SMC-analog are indicated with a white dashed arrow. These visualizations highlight the emergence of stripped gas and stars as the interaction progresses.

This interaction is reminiscent of current models of formation of a tidally-stripped Magellanic Stream,
\begin{figure*}[!htp]
\centering
\includegraphics[width=0.91\linewidth]{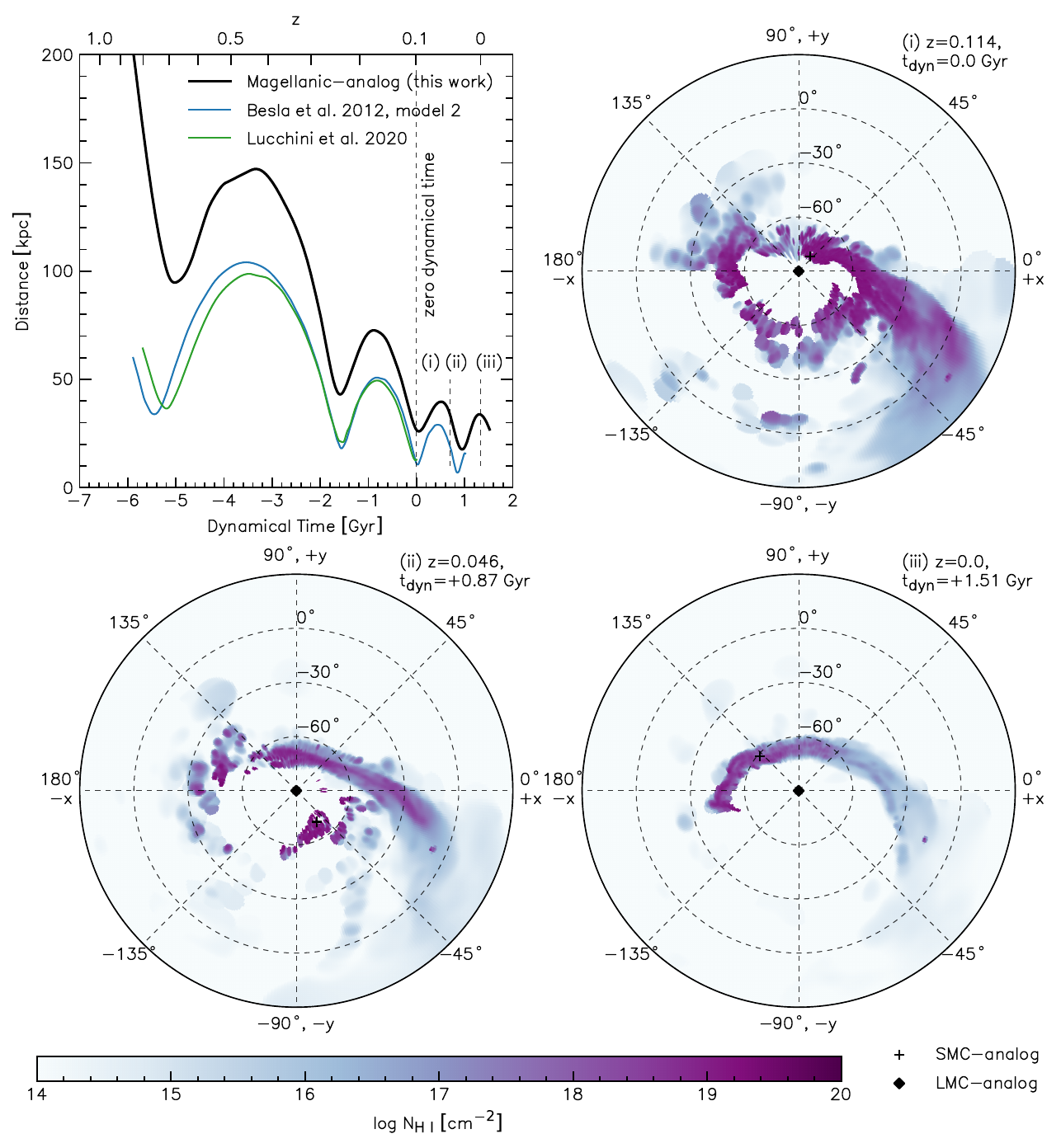}
\caption{Dynamics of the Magellanic-analog system. [\textit{upper left panel}] The distance between the LMC- and SMC-analog in our simulation as a function of dynamical time $t_{\text{dyn}}$ (as defined in \S\ref{sec:dynamical_time}) and redshift is shown as the solid black line. The same data from \citet{Besla+2012}, model 2 
isolated
without the influence of the MW, and likewise from \citet{Lucchini+2020} are shown as the solid blue and green lines, respectively. Note: The \citet{Besla+2012} and \citet{Lucchini+2020} curves correspond to isolated simulations, and the axis denoting redshift only applies to the curve of our Magellanic-analog dwarf pair. [\textit{remaining panels}] Stereoscopic projections of the tidally-stripped gaseous stream  colored by \ion{H}{1} column density at: $(i)$ zero dynamical time $t_{\rm dyn}=0.0$ Gyr, $(ii)$ $t_{\rm dyn}=+0.87$ Gyr, $(iii)$ $t_{\rm dyn}=+1.33$ Gyr. $N_{\rm HI}$ values were calculated from the frame of reference of an observer placed at $(x, y, z)=(0, 0,+55\text{ kpc})$ above the plane of the LMC-analog. Thus, the image shows the synthesized viewpoint of an observer facing downwards (in the $-z$ direction) towards the Magellanic-pair analog. The SMC- and LMC-analogs are shown as a black $``+"$ and black diamond, respectively; the SMC-analog orbits in the counterclockwise direction from this perspective.
}
\label{fig:dynamics}
\end{figure*}
including \citet{Besla+2012}'s second model and \citet{pardy+2018}'s 9:1 model, which was subsequently adopted by \citet{Lucchini+2020}. 
The separation distance between the two dwarfs as a function of dynamical time (in which the stopping time of interactions of previous models is mapped appropriately to our pair of dwarfs) can be found in Figure \ref{fig:dynamics}, \textit{upper left panel}.
These properties of the interacting dwarf pair, combined with the interaction's similarities to existing paradigm LMC-SMC interaction histories \citep{Besla+2012} 
provides an opportunity to explore the properties and dynamics of a Magellanic system-analog prior to its interaction with the Milky Way halo, where they are implied to be more massive than present day \citep{pardy+2018}.
The properties of the system are summarized in Table \ref{tab:properties}. 

Furthermore, with regards to the remaining massive dwarf halos within the three highest resolution \textsc{hestia} runs, there exist 28 dwarf galaxies of approximately LMC-mass. Among these, four possess bound satellites of SMC-mass that have completed at least one orbit around their host within the total simulation time (age of the universe). Of these four dwarfs, two form tidally-stripped neutral streams of gas: one of which is the topic Magellanic-analog system, and the other of which is substantially more diffuse\footnote{The volumetric number density $n_{\rm HI}$ of the remaining tidally-stripped neutral stream of gas is $\sim 1/10$ that of the Magellanic stream-analog discussed in this paper, significantly more diffuse than the Magellanic Stream, and so we decided to omit its discussion.}. Therefore, we find a fraction of $1/7$ of LMC-mass galaxies possessing an SMC-counterpart in our simulation suite, matching decently well with the expected fraction of $1/5$ from mock catalogs constrained by observations \citep{rodriguez+2013}.

\subsection{Coronal Gas Properties in a Massive Dwarf Halo}
The LMC-analog in our Magellanic-analog system exhibits a well-defined, coronal gas envelope that extends approximately to its virial radius, as can be seen in Figure \ref{fig:images}, \textit{center column}.
 The density of this corona follows a declining power-law relation, with $n_{\rm{H}} \sim 10^{-5}$ cm$^{-3}$ near the disk and decreasing to $\sim 10^{-6}$ cm$^{-3}$ at larger radii. 
 The temperature of the corona, $T\sim 2.6\times10^5$ K, is in agreement with expectations from the virial theorem for a halo of this mass \citep{mo+2010} and is consistent with constraints from observational studies of the LMC’s coronal gas \citep{Krishnarao+2022}\footnote{
Direct comparison with Krishnarao et al. (2022) is not straightforward, as their observations reflect the properties of an LMC corona influenced by the Milky Way's CGM and gravitational potential. Our analog system is dynamically isolated prior to infall, representing a different evolutionary stage.}

\begin{figure*}
\includegraphics[width=\linewidth]{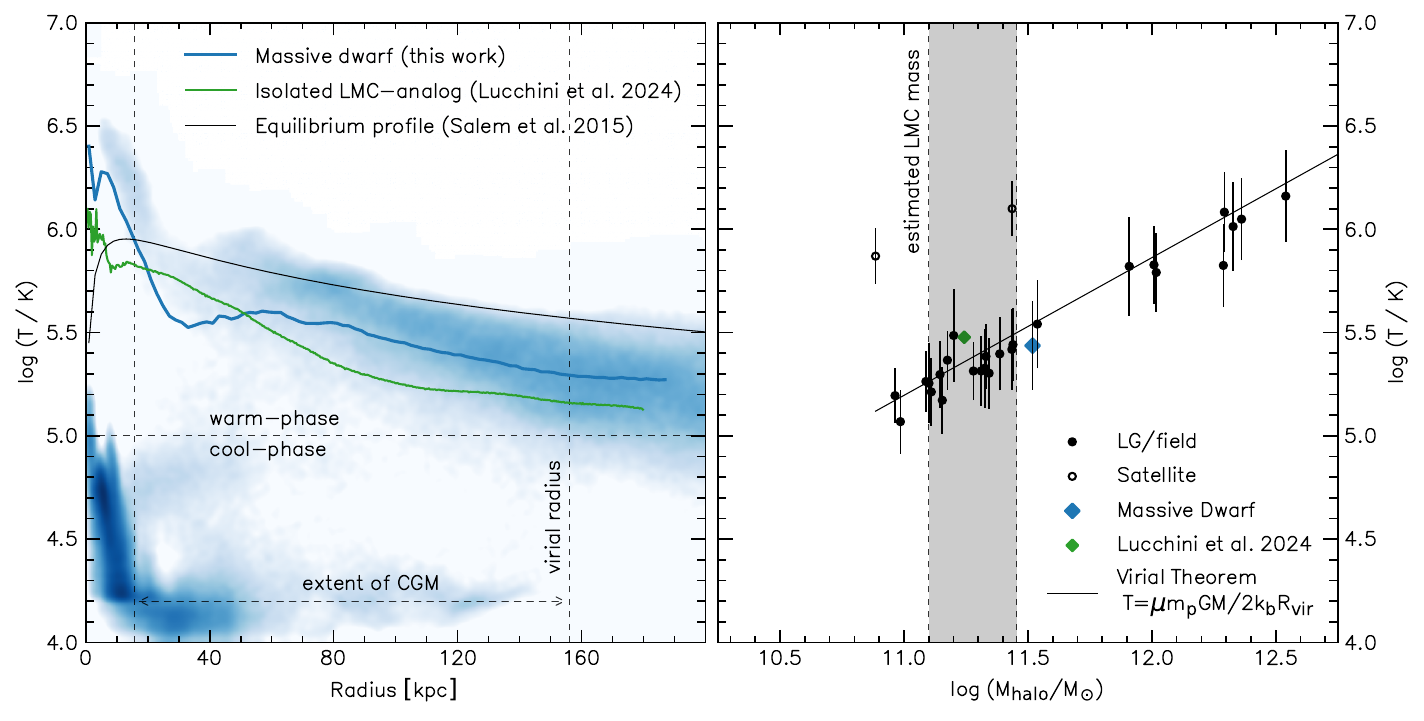}
\caption{Temperature profile of the coronae at zero dynamical time ($z\sim0.1$). [\textit{left panel}] Heatmap of the radial distribution of the temperature of gas cells of the CGM of the LMC-analog (darker blue corresponding to higher phase space density). The mass-weighted average temperature of \textit{strictly the warm phase} of the CGM at each radial bin is shown as a solid blue line. The corresponding temperature profile for the slightly less-massive isolated LMC-analog from \citet{Lucchini+2024} is shown as a solid green line. The expected analytic profile for a virialized halo with a mass of our massive dwarf is shown as a solid black line.
[\textit{right panel}] The temperature of coronae as defined in \S\ref{sec:corona_characterization} as a function of halo mass. The solid points correspond to isolated halos (members of the Local Group-analog or field dwarfs), and hollow points correspond to those within the virial radius of a larger host (satellites). Our massive dwarf is represented with a blue diamond, and the fiducial model of \citet{Lucchini+2024} is represented as a green diamond. Vertical error bars denote the $1\sigma$ spread in the estimated temperature of the corona for a given dwarf. The virial temperature as a function of halo mass is shown with a dashed black line. The estimated LMC virial mass from globular clusters \citep{watkins+2024} is denoted with the vertical gray band.
}
\label{fig:gas}
\end{figure*}

A comparison of the simulated temperature profile at zero dynamical time with previous theoretical models is shown in Figure \ref{fig:gas} (\textit{left panel}). When accounting for the slightly differing halo masses, our results closely match the idealized isolated simulations and LMC corona of \citet{Lucchini+2024, Lucchini+2020}, as well as the hydrostatic equilibrium model of \citet{salem+2015} for the CGM of a similar-sized halo. 

The multiphase structure of the CGM is displayed in the heatmap of gas cells in Figure \ref{fig:gas}, \textit{left panel}. The region in the bottom-left illustrates the cooler clumps and streams of gas, as well as the cool gaseous disk; the horizontal band of gas stretching from 40 kpc to $R_{\rm vir}$ at $T \sim 3\times 10^5$ K denotes the corona. Furthermore, the vertical band of gas around $R\sim40$ kpc includes interfaces between the cool and warm phases of the CGM, as well as interfaces between the ionized CGM and neutral gas stream. The definitions of the warm and cool phases of the CGM, as well as extent of the CGM, as defined in \S\ref{sec:corona_characterization}, is also denoted in the figure.

The temperature profile of the warm phase of the CGM of our massive dwarf matches well with that of the isolated LMC-analog from \citet{Lucchini+2020}, once adjusted due to the slightly differing masses between the two dwarfs. This is particularly true at higher radii, where gas is expected to be less dense, and therefore the warm phase of the CGM dominates. Despite our corona profile tracing exclusively the warm phase, below $R\sim0.5 R_{\rm vir}$, the interfaces between cool and warm phases of the CGM pull the average temperature of the warm gas down. This multiphase inner CGM was not present in \citet{Lucchini+2020}'s LMC-analog due to its construction as a simple halo of warm-hot gas. That corona was also unstable, likely due to the absence of a regulatory cool phase of the CGM, hence the lack of trough in its temperature profile at low radii. Our self-consistent and stable corona is formed from the virialization of the LMC-analog's halo, a phenomena made possible by the cosmological nature of the \textsc{Hestia} simulations.

\subsection{Formation and Evolution of the Neutral Gas Stream}

Furthermore, tidal interactions between the two dwarf galaxies lead to the formation of a coherent neutral gas stream extending over 150 kpc. The LMC-analog interacts with the SMC-analog, stripping neutral gas from the latter and producing a stream, consistent with theoretical models of dwarf–dwarf interactions \citep{DonghiaNat09}.

\begin{deluxetable*}{lllrrrrrr}[t]
\tabletypesize{\scriptsize}
\tablewidth{0pt} 
\tablehead{
\colhead{} & \colhead{$z$} & \colhead{Object} & \colhead{$\log (M_{\text{halo}} / M_{\odot})$}& \colhead{$\log (M_{\text{gas}} / M_{\odot})$} & \colhead{$\log (M_{*}/M_{\odot})$} & \colhead{$\log (n_{\text{H}} / \rm{cm}^{-3})$} & $\log (T/K)$ & \colhead{References} \\
}
\startdata 
\multirow{12}{4em}{\textsc{Hestia}} & \multirow{4}{3em}{0.5} & LMC-analog & 11.49 & 10.61 & 9.98 & - & - & \\
&  & SMC-analog & 10.09 & 9.42 & 8.61 & - & - & \\
&   & \ion{H}{2} Corona & - & 9.57 & - & $-5.71(37)$ & $5.42(25)$ & \\
&   & \ion{H}{1} Stream & - & - & - & - & - & \\
\cline{2-8}
& \multirow{4}{3em}{0.1} & LMC-analog & 11.52 & 10.61 & 10.14 & - & - & \\
&  & SMC-analog & 9.47 & 8.68 & 8.41 & - & - & \\
&   & \ion{H}{2} Corona & - & 9.88 & - & $-5.77(23)$ & $5.44(21)$ & \\
&  & \ion{H}{1} Stream & - & 8.32 & - & $-3.33(51)$ & $4.07(19)$ & \\
\cline{2-8}
& \multirow{4}{3em}{0.0} & LMC-analog & 11.52 & 10.57 & 10.21 & - & - & \\
&  & SMC-analog & 8.65 & - & 7.91 & - & - & \\
&   & \ion{H}{2} Corona & - & 9.32 & - & $-5.79(20)$ & $5.42(18)$ & \\
&  & \ion{H}{1} Stream & - & 7.57 & - & $-3.41(29)$ & $4.08(08)$ & \\
\hline
\multirow{4}{4em}{Observations} &  & LMC & 11.3(1) & 8.64(1) & 9.4 & - & - & [1,2,3]\\
&  & SMC & $\geq9.8$ & $8.60(1)$ & 8.5 & - & -& [4,2,5]\\
&   & \ion{H}{2} Corona & - & $9.10(2)$ & - & - & $\sim 5.5$ & [6,6]\\
&  & \ion{H}{1} Stream & - & 8.69 & - & $-2.0(2)$ & $< 4.2$ & [2,7,8]\\
\enddata
\caption{Properties of the various components in the Magellanic-analog dwarf pair following virialization of the halo of the massive dwarf ($z\sim0.5$), at zero dynamical time ($z\sim 0.1$), and at present day. References: [1] \cite{shipp+2021}; [2] \cite{bruns+2005}, \textit{assuming a distance of $50$ kpc, $60$ kpc, and $55$ kpc for LMC, SMC, and Stream, respectively}; [3] \cite{vanDerMarel+2008}; [4] \cite{bekki+2009}, [5] \cite{snez+2004}; [6] \cite{Krishnarao+2022}; [7] \cite{fox+2014}, \textit{50 kpc model} [8] \cite{hsu+2011}.}
\label{tab:properties}
\end{deluxetable*} 

This \ion{H}{1}-rich structure primarily originates from the lower-mass companion and contains a total neutral hydrogen mass of $M_{\rm HI}= 2.09\times10^8\text{ M}_{\odot}$ immediately following formation (then decreasing to $M_{\rm HI}= 3.72\times10^7\text{ M}_{\odot}$ after $\sim$1.5 Gyr), similar to the observed Magellanic Stream, with a distance assumption of $\sim 55$ kpc \citep{bruns+2005}. 
\ion{H}{1} column densities of gas associated with the SMC-analog, as measured by an observer at this distance, at various snapshots following stream formation ($t_{\text{dyn}}=0.0 \text{ Gyr}$), are shown in Figure \ref{fig:dynamics}, (\textit{upper right and lower panels}). The column densities of our stream-analog lie primarily between $10^{17} \text{ cm}^{-2}<N_{\rm HI} < 10^{19} \text{ cm}^{-3}$, consistent with observations from the Parkes \ion{H}{1} survey, $10^{18} \text{ cm}^{-2} <N_{\rm HI}^{\rm Parkes} < 10^{19} \text{ cm}^{-3}$ \citep{bruns+2005}. 
We note that our synthetic column densities are sensitive to the choice of reference frame. The inclusion of a Milky Way potential would likely elongate the stream, resulting in lower column densities due to increased path length and dispersion.

A key question regarding the Magellanic Stream is its long-term survival against interaction with the surrounding CGM. In our simulation, the neutral gas stream retains a relatively low velocity with respect to the coronal gas, $v_{\rm rel} \sim 120$ km s$^{-1}$, which minimizes the impact of ram-pressure stripping and instabilities such as Kelvin-Helmholtz and Rayleigh-Taylor instabilities \citep{murray+1993, bland-hawthorn+2007}. Over a timescale of $\sim 1.5$ Gyr, we find only a marginal decline in the total \ion{H}{1} mass of the stream, suggesting that a neutral gas stream embedded in a warm corona under these conditions can remain stable for extended periods of time. This result is consistent with prior theoretical work indicating that neutral structures can survive within a warm-hot CGM if the relative velocity remains sufficiently low \citep{Murali+2000, tepper-garcia+2019}. It also reinforces observational findings that the Magellanic Stream might have persisted for a few Gyr despite interactions with the Milky Way’s CGM \citep{nidever+2008, fox+2014}.

\subsection{Ubiquity of Coronal Gas surrounding Massive Dwarf Galaxies}

Beyond the specific Magellanic-analog system, we investigate the presence of coronal gas in all massive halos within the \textsc{Hestia} simulations. We find that every halo with $\gtrsim 10^{11}$ M$_{\odot}$ hosts a warm-hot circumgalactic corona, with a mass-temperature relationship following the expectations of the virial theorem (Figure \ref{fig:gas}, \textit{right panel}). The remaining isolated dwarfs exhibit CGM temperatures of $T \sim 3 \times 10^5$ K, consistent with theoretical predictions for hydrostatic equilibrium \citep{mo+2010}. 
However, some massive dwarfs residing within the virial radius of a larger halo exhibit higher temperatures, likely due to the interactions with the even warmer CGM of the more massive host.
This suggests that environmental effects—such as pressure confinement and hydrodynamic mixing—may alter the thermal properties of dwarf galaxy coronae, a hypothesis that should be tested in future observational studies.

Our result is consistent with previous simulation-driven studies \citep{jahn+2022}, as well as recent UV-absorption-line surveys which detect coronal gas surrounding massive dwarf galaxies \citep{bordoloi+2014, johnson+2017, zheng+2024, nishant+2024, piacitelli+2025}, independent of their location within a group or cluster environment. The presence of such coronae likely also has significant consequences for gas accretion and star formation regulation in dwarf galaxies, similar to the role of coronal gas in Milky Way mass halos \citep{tumlinson+2017, stern+2021}.



\section{Discussion}
\label{sec:discussion}

The interacting dwarf galaxy pair identified in our simulation provides a rare opportunity to study the pre-infall evolution of a Magellanic-analog system within a fully cosmological environment. The dynamical history, gas stripping, and circumgalactic structures of this system offer a direct point of comparison with both idealized simulations and observational data of the Magellanic Clouds.

This work presents, to our knowledge, the first identification of a pre-infall LMC–SMC analog pair within a constrained cosmological volume that reproduces both a warm circumgalactic corona and the formation of a neutral tidal stream. While previous studies, such as \citet{jahn+2022}, have identified Magellanic-like coronae in zoom-in simulations, our analog allows us to investigate the coupled evolution of the coronal medium and gas stripping during repeated interactions, providing new insight into the timing, morphology, and survival of Magellanic-like gaseous structures prior to infall. This complements earlier work such as \citet{Lucchini+2020, Lucchini+2024}, which used the \textsc{GIZMO} simulation code, and reinforces the idea that massive dwarfs naturally host coronae due to virial shock heating.

The orbital history of the SMC-analog closely mirrors expectations from previous models of the Magellanic Stream's formation. 
Our choice of using a dynamical time defined relative to the third pericenter passage—corresponding to maximal gas loss in prior models—allows for direct alignment of our simulation’s timeline with key physical events observed in the Magellanic System.
Over a period of $\sim 6$ Gyr, the two dwarfs complete four pericentric passages, with pericenter distances declining to $\sim$15 kpc. 
The neutral stream begins to form after the third passage, when tidal forces strip the outer layers of the SMC-analog's gas envelope. Despite strong interactions, the SMC-analog remains a distinct bound system through to $z=0$. The neutral gas stream's formation thus follows the classical tidal stripping scenario described in \citet{Besla+2012, Lucchini+2020}.

In addition to the trailing stream, we identify structures that resemble both the Magellanic Bridge and the Leading Arm. The bridge, as labeled in the second stellar surface density panel in Figure \ref{fig:images}, appears as a filament of both stars and neutral gas connecting the two dwarfs during their intermediate encounters. It likely consists of mixed material, including gas chemically linked to the SMC-analog and stars originating from both galaxies. This structure is consistent with the observed Magellanic Bridge in position, density, and timing \citep{putnam+2003}. A diffuse feature preceding the stream composed of clumpy neutral and ionized material, analogous to the Leading Arm \citep{putnam+1998, fox+2018, for+2013}, forms shortly after the trailing stream emerges, and is visible in the upper right and lower left panels of Figure~\ref{fig:dynamics}. However, this feature dissipates within $\sim$600 Myr, likely due to evaporation\footnote{
Here, we refer to “evaporation” as the gradual thermal and hydrodynamic dissolution of neutral gas into the surrounding warm halo. While our simulation does not explicitly resolve microphysical thermal conduction or turbulent mixing, the behavior we observe —disruption of cool clumps— is consistent with these processes at sub-grid scale. The Leading Arm analog is more diffuse, lower in column density, and less shielded by the host potential than the trailing stream, making it more vulnerable to disruption by the ambient corona. In contrast, the trailing stream is denser, more compact, and originates closer to the LMC-analog disk, contributing to its longer survival time.}
within the LMC-analog's coronal environment. Its short survival time, even in the absence of external stripping from a Milky Way halo, suggests that the real Leading Arm may be shaped or preserved by ram pressure from the Milky Way’s CGM and the LMC coronal gas, or place timing constraints on formation of the Magellanic stream. Nonetheless, our results thus place an upper limit on the Leading Arm's survivability in a purely dwarf–dwarf interaction context.

While the total \ion {H}{2} mass of the LMC-analog's corona agrees well with observations, we do not observe significant stripping of the LMC-analog's coronal gas into the stream, as present in \citet{Lucchini+2020}. This is likely due to the limitation of our model, more specifically the absence of a Milky Way potential and CGM in our setup. While some ionized gas interfaces exist between the stream and the warm corona, they do not constitute a dominant component. This reinforces the idea that the observed ionized portion of the Magellanic Stream requires environmental processing, such as ram-pressure stripping, after infall. As emphasized by \citet{fox+2014}, the ionized fraction is substantial in the observed Stream, likely originating from the LMC's corona during interaction with the Milky Way \citep{Lucchini+2020}.

In addition to these tidal gaseous features, we identify a very faint ($\Sigma \sim 10^5$ M$_{\odot}/$kpc$^2$, corresponding to a V-band surface brightness of $\sim28$ mag/arcsec$^2$ when assuming a distance of 55 kpc) tidal stellar stream, which forms immediately following the formation of the \ion{H}{1} stream, dispersing to an even lower surface density through to $z=0$. This stellar stream, labeled in the lowest stellar surface density plot in Figure \ref{fig:images}, precedes the gaseous stream, likely due to the necessarily stronger tidal forces required to eject the star particles from their deeper potential well. We see a similar phenomena in the other tidally stripped stream of an SMC-mass galaxy by an LMC-mass halo in our simulation suite (as mentioned in \S\ref{sec:indentification_of_system}). This strongly suggests that the Magellanic stellar stream, if it exists, may not necessarily be spatially coexistent with the \ion{H}{1} stream. Recent observations are consistent with this; they indicate stars of SMC origin are identified in the subdominant strand of the Magellanic stream, spatially offset and distinct from the dominant strand, which contains much higher \ion{H}{1} column densities \citep{zaritsky+2025, chandra+2023}.


The coronal gas envelope surrounding the LMC-analog in our simulation exhibits a stable, multiphase structure extending to the virial radius. Its temperature and density profiles are consistent with theoretical expectations from the virial theorem and hydrostatic equilibrium \citep{salem+2015}. Unlike prior work in idealized simulations, we resolve the corona's structure in a cosmological setting, where the halo forms naturally from virialized accretion. 

In the \textsc{Hestia} suite, all massive dwarfs with $M_{\rm halo} \gtrsim 10^{11}\mathrm{M}_\odot$ form such coronae, indicating this is a robust outcome of cosmological structure formation. A comparison to observational constraints, such as those in \citet{Krishnarao+2022}, is not straightforward because the observed Magellanic corona has already experienced infall and interaction with the Milky Way CGM, whereas our system is dynamically isolated. We find that cool clumps of HI do not persist longer than $\sim$600 Myr within the warm halo, a limit potentially relevant for interpreting the Leading Arm's longevity.


Our findings highlight the importance of environment and pre-processing in shaping the CGM and gaseous structures of dwarf galaxies. The longevity of the trailing stream, the fragility of the leading feature, and the composition of the bridge all suggest that dwarf–dwarf interactions imprint observable signatures before infall, but that additional environmental mechanisms (e.g., ram pressure) are likely required to reproduce the full complexity of the Magellanic System. In future work, we will introduce a Milky Way-mass halo into the evolution of this analog system to directly investigate the impact of infall, gas stripping, and stream survival in a realistic Local Group context.

\section{Conclusion}\label{sec:conclusion}

We present the first identification of a pre-infall LMC–SMC analog pair within the \textsc{Hestia} constrained cosmological simulation suite. This system provides a rare laboratory for exploring the coupled evolution of dwarf galaxy interactions, circumgalactic coronae, and neutral gas streams before environmental effects from a Milky Way-mass halo come into effect.

The LMC-analog hosts a stable, warm, virialized corona, and the pair undergoes repeated tidal interactions over $\sim$6 Gyr. These interactions result in the stripping of the outer gas from the SMC-analog, producing a $\sim$150 kpc-long neutral stream with an \ion{H}{1} mass comparable to that estimated of the observed Magellanic Stream. Despite the strength of the interactions, the SMC-analog remains a distinct, bound system through to $z = 0$, consistent with classical tidal stripping scenarios. 

We also identify a bridge-like structure composed of both stars and gas, and a short-lived leading feature analogous to the Leading Arm. The latter dissipates within $\sim$600 Myr, likely due to evaporation in the LMC-analog’s corona, placing an upper bound on the survival time of such structures without ram pressure from a host halo. Furthermore, we identify a stellar component to the tidal stream, out of phase from the \ion{H}{1} stream, likely due to the stellar disk's deeper location within the halo potential well than the gaseous disk, and the dwarfs' repeated close interactions. This strongly implies that the Magellanic stellar stream, if it exists, may not be spatially coexistent to the dominant \ion{H}{1} stream.

The LMC-analog's ionized CGM is not yet stripped into the trailing stream due to the absence of a Milky Way-analog and its ram-pressure striping effects; however, both the neutral and ionized gas at their necessary masses are present in our pre-infall system.
These results suggest that the neutral portion of the Stream and related features can form pre-infall, but their long-term survival and ionized morphology require further environmental processing.
Our findings here provide an essential pre-infall baseline for interpreting the observed Magellanic System, may set expectations for future observations of dwarf galaxies in the Local Group and serve as a theoretical benchmark for upcoming facilities like JWST, SKA, and the Roman Space Telescope.

\section{Acknowledgments}
R.C. thanks the \textit{National Space Grant College and Fellowship Program} and the \textit{Wisconsin Space Grant Consortium} for their support (RFP24\_5-0, RFP25\_7-0,). E.D. acknowledges the grant, HST-AR-17053.004-A.

%

\vspace{5mm}
\facilities{\textit{erebos} computing cluster, Leibniz-Institut f\"ur Astrophysik Potsdam (AIP)}


\software{Numpy \citep{vanDerWalt+2011}, SciPy \citep{virtanen+2020}, Matplotlib \citep{hunter+2007}, Astropy \citep{astropy:2018}}

\bibliography{references}{}
\bibliographystyle{aasjournal}



\end{document}